# Risk as Challenge: A Dual System Stochastic Model for Binary Choice Behavior

Samuel Shye♣      Ido Haber♠

## Abstract

*Challenge Theory* (*CT*), a new approach to decision under risk departs significantly from expected utility, and is based on firmly psychological, rather than economic, assumptions. The paper demonstrates that a purely cognitive-psychological paradigm for decision under risk can yield excellent predictions, comparable to those attained by more complex economic or psychological models that remain attached to conventional economic constructs and assumptions. The study presents a new model for predicting the popularity of choices made in binary risk problems.

A CT-based regression model is tested on data gathered from 126 respondents who indicated their preferences with respect to 44 choice problems. Results support CT's central hypothesis, strongly associating between the *Challenge Index* (*CI*) attributable to every binary risk problem, and the observed popularity of the bold prospect in that problem (with $r=-0.92$ and $r=-0.93$ for gains and for losses, respectively). The novelty of the CT perspective as a new paradigm is illuminated by its simple, single-index (*CI*) representation of psychological effects proposed by Prospect Theory for describing choice behavior (certainty effect, reflection effect, overweighting small probabilities and loss aversion).

## 1. Introduction

In the Expected Utility Theory (EUT) the utility of an uncertain prospect is the sum of the utilities of the outcomes, each weighted by its probability. In choosing between two or more prospects, the rational decision-maker should prefer a prospect that maximizes utility (von Neumann and Morgenstern, 1947; Savage, 1954). Given a decision problem whose prospects have different utilities, a unique prospect should be chosen by different people as well as by the same person on different occasions.

Systematic violations of the EUT observed in human behavior have led to the development of descriptive theories, that aim to predict people's actual decision behavior. Such descriptive theories tend to adopt EUT's conceptual framework as a point of departure and proceed to explain observed violations of EUT by introducing modifications to this basic theory. Thus, Prospect Theory (PT) (Kahneman &

♣ Department of Psychology and PEP Program (Philosophy, Economics & Political Science), The Hebrew University of Jerusalem; and The Van Leer Jerusalem Institute. Email: samuel.shye@mail.huji.ac.il
ORCID: 0000-0001-6277-5969

♠ Technion – Israel Institute of Technology, Haifa; and Szold Institute, Jerusalem. Email: ido.haber@campus.technion.ac.il.

Research was partially supported by The National Mentoring Program, Szold Institute, Jerusalem

Tversky, 1979; Tversky & Kahneman, 1992), arguably the most influential descriptive model, accounts for observed violations of EUT axioms by proposing functions that assign effective values to gains and to losses (rather than to the total assets); and effective weights to the given probabilities. A different modification of EUT is proposed, e.g., by Regret Theory (Loomes & Sugden, 1982, 1987): a prospect's utility-function is fashioned to incorporate feelings of regret and rejoicing anticipated by the decision maker. The decision maker aims then to maximize the expected *modified utility* function.

PT and similar descriptive models, referred to as *deterministic*, follow in EUT's footsteps in their aim to predict the *one* prospect that is preferable to all others, where the empirical confirmation of the prediction is embodied in the *modal* prospect, the one preferred by most respondents in a sample of decision makers. Striving for a unique solution may be natural for an economic-prescriptive theory concerned with profit optimization but is less natural for a genuine *descriptive* theory, concerned with predicting people's actual behavior. From such a theory one would expect a richer characterization of choice behavior, such as people's *inclination* to prefer each of the prospects offered, as reflected in the distribution of responses to a given risk problem. Indeed, *stochastic* decision theories have been proposed (for example, Busemeyer & Townsand, 1993; Erev et al., 2002) that aim to predict the popularity of a particular prospect, notably in binary risk problems.

A distinct birthmark inherited from EUT by many descriptive theories is the assignment of a value $V(f)$ to every prospect $f$, such that prospect $f$ is preferred to prospect $g$ iff $V(f) > V(g)$. And, in line with EUT, $V(f)$ is computed as the sum of the outcome utilities, each weighted by a transformation of corresponding probability (Kahneman & Tversky, 1979) or by decision weights that derive from transforming the entire cumulative distribution function (Tversky & Kahneman, 1992). Thus, predictions of descriptive theories assume (tacitly at least) that a decision maker faced with two or more risky prospects performs in effect the following *mental* operations:

(a) *Multiplying* the outcome-value by the outcome probability-weight to obtain the expected utility for each outcome;

(b) *Summing up* the expected utilities of all outcomes of a prospect, to obtain the prospect's value $V(f)$;

(c) *Comparing* the overall utilities of all prospects offered, identifying one of highest value.

The essential structure of the prospect-utility assessments (steps (a) and (b)) is clearly illustrated by the weighted-sum *form* of the value $V(f)$ assigned to each prospect $f$:

$$V(f) = \sum [v(x_i) \cdot w(p_i)]$$

(See, for example, Tversky & Kahneman, 1992, Eq. (2); Loomes & Sugden 1982, Eq. (2); Erev et al., 2010, Eq. (1)).

Having determined the value $V(f)$ for each of the available prospects, step (c), of comparing every pair of prospects, amounts to computing the difference between their values (for example, Loomes & Sugden, 1982, eq. (4)) or a monotone function of that difference (Erev et al., 2010, eq. (4)). This is true for stochastic theories as well for deterministic theories. (See, for example, Busemeyer & Townsand, 1993; Erev et al., 2002).

While the three operations (a)-(c) may well be performed by adequately schooled people acting within a professional setting, they are not likely to be mentally carried out by people (schooled or not) in daily intuitive decision making. Evidently, descriptive theories have not substantiated or rationalized these mental operations to be part of any psychological modeling of decision making under risk. Rather, most descriptive theories began by accepting EUT and proceeded to modify it just as necessary in order to account for observed violations of its axioms and prescriptions.

Adherence to EUT conceptual framework, including the weighted-sum criterion traditionally constrained the development of a genuine descriptive theory for choice under risk. Hence psychological theories that focus on the decision process rather than on its consequences, have emerged. See Weber & Johnson (2009) for a review of choice phenomena research that hinges on psychological -- cognitive and affective -- processes. Among these, processes that rely on the dual system approach (Stanovich & West, 2000; Kahneman, 2011) have a special appeal in as much as they attempt to integrate two distinct kinds of processes that directly concern stimuli evaluation and decision-making: the automatic system and the analytic system. Indeed, dual processing models for decision making have been employed for explaining decision behavior (e.g., Kahneman, 2003; Sanfey et al., 2006).

The *automatic system* (also referred to as System 1 or Type 1 processing) is autonomic: The execution of System 1 processing is involuntary and typically rapid and unconscious (Stanovich, 2011). System 1 processing has been described as associative and heuristic (e.g., Sloman, 1996; Evans, 1984; 1989); as relying on intuitive cognition (Hammond, 1996); as tacit thought processes (Evans & Over, 1996) including tacit inferences (Johnson-Laird, 1983); as an automatic activation (Posner & Snyder, 1975); and as evolutionary old (Evans, 2008). In contrast with the automatic system, processing of the *analytic system* (also referred to as System 2 or Type 2 processing) is deliberate and slow, involving controlled analytic intelligence. It has also been described as a rule-based system (e.g., Sloman, 1996); as relying on analytical cognition (Hammond, 1996); as an explicit thought process (Evans & Over, 1996) including explicit inferences (Johnson-Laird, 1983); as a conscious processing system (Posner & Snyder, 1975) and as evolutionary new (Evans, 2008).

The automatic system, like the analytic system, is characterized by an emphasis on the cognitive behavior mode (cf. Evans, 2008. Note also its description in terms of intuitive *cognition*; *thought* processes; and *inferences*). Yet, free from a higher control system, it also incorporates beliefs and affective behaviors, shaped by psychological and social evolutionary processes, as fast, effort-saving ways for heuristically coping with adaptive problems (Finucane et al., 2000; Loewenstein et al., 2001). However, under the umbrella of the dual system approach to decision-making, the two systems are often interpreted as representing the very distinction between the affective and the cognitive modes of decision-making behavior (see e.g., Rottenstreich & Hsee, 2001; Hsee & Rottenstreich, 2004). An application of the affective-cognitive duality to decision making is proposed by Mukherjee's (2010) model depicting the decision process as a convex combination of affective and cognitive processes, each marked by characteristic shapes of the value function and of the weighting function.

In this article, we develop a new theory, called the Challenge Theory (CT), that examines decisions under risk as a behavioral phenomenon in and by itself, free of any preconception of which decisions are rational or otherwise desirable; a domain of investigation in which stable empirical patterns, and ultimately laws, are to be discovered, leading to predictions in this domain. In so doing we believe we follow the common practice in theory-oriented empirical sciences.

Challenge Theory is a dual system theory in that it acknowledges that two kinds of processes, two systems, participate in decision making: the fast (automatic) and the slow (analytic). While formulating the two systems in essentially cognitive terms (à la Stanovich, e.g.), CT does not preclude the possibility that decision maker's feelings, actual or anticipated, play some role in *both* systems: in the automatic system, as a fast way to respond; and in the analytic system as an object (among others) to be considered in the deliberate calculations. Furthermore, CT models the systems as working sequentially, with the automatic processing reaching a fast, initial decision, followed by the analytic processing, acting in a supervisory role: affirming or modifying the initial decision. This is in contrast with Mukherjee's (2010) dual system model, where two processing systems, the one defined as affective and the other as cognitive, are described as operating in parallel.

The plan of this article is as follows. Section 2 presents a reconstruction of the decision maker's cognitive process in the face of a binary choice problem, culminating in a formula that assesses the degree of difficulty, or *challenge-level*, experienced by the decision maker in abandoning the option defined here as the *default option* and choosing the option defined here as the *bold option*. Section 3 presents and tests the General Hypothesis of Challenge Theory, namely, that the *challenge level*, attributable to every binary choice problem, predicts people's inclination (probability) to choose the bold option in that problem. The possibility afforded by the Challenge Theory to explore individual differences is illustrated in Section 4. The discussion in Section 5 elaborates on the specifications, significance and implications of Challenge Theory as a novel paradigm. It also elaborates on CT's integrated representation of psychological effects proposed by Prospect Theory, as well as on some general questions – the role of extant wealth in gambling and the gain-loss asymmetry.

## 2. Decision-making as a mental process: a reconstruction

We are concerned with choice problems where the decision-maker must choose one of $k+1$ prospects, $(x_0, p_0), (x_1, p_1) \ldots (x_k, p_k)$, wherein the outcome of each is a monetary gain, $x_i$ ($0<x_i<\infty$) with a known probability $p_i$ ($0<p_i\leq1$), $i=0,1,\ldots,k$ and 0 with probability $(1-p_i)$. We assume $p_0>p_1>p_2>\ldots>p_k$ and $0<x_0<x_1<\ldots<x_k$.

Alternatively, the outcome of each of the $k+1$ prospects is a monetary loss, $x_i$ ($-\infty<x_i<0$) with a known probability $p_i$ ($0<p_i\leq1$), $i=0, 1,\ldots,k$ and 0 with probability $(1-p_i)$. Here, $x_i$ being all negative, we assume $p_0>p_1>p_2>\ldots>p_k$ and $0>x_0>x_1>\ldots>x_k$.

A choice problem with these characteristics will be called a *simple decision problem*. In this article we consider simple *binary* decision problems, those with $k=1$. This class of problems has been used extensively in previous studies and simulates many common problems encountered in real-life situations. The ideas developed here can be extended to decision problems with any $k$. Our aim is to develop a theory, as simple and purposeful as possible, that would explain and enable predictions of people's inclination (probability) to prefer one of the prospects in a binary decision problem over the other.

*2.1 The case of gains*

Let us consider first gain problems, i.e., decision problems with $x_1 > x_0 > 0$. The decision-maker is faced with two options $(x_0, p_0)$ and $(x_1, p_1)$, each containing two numerical parameters, the (positive) outcome, $x_i$, and the probability of receiving it, $p_i$. The total information input consists therefore of four numerical values to be processed mentally in order to reach a decision. The decision-maker's initial impression of the problem is governed by his automatic mental processing system (*System 1 or Type 1 processing,* (Stanovich & West, 2000; Stanovich 2011; cf. also Kahneman, 2011), which, we claim, stresses the probabilities rather than the outcomes. Hence, we assume that a person's initial attraction, or default preference, would be to $(x_0, p_0)$, the prospect that promises the highest probability for *some* gain. This assumption is based on an extension of the certainty effect (Kahneman & Tversky, 1979) which posits the preference of the sure outcome over the unsure outcome, to the more general assertion that posits the preference of the *surer* (more probable) outcome over the less sure outcome. But there is a more substantive rationale for $(x_0, p_0)$ being the heuristic choice of the automatic processing system: this system, by its very nature, concerns the most vital organismic needs, including the survival of the organism. If gaining *some* amount, however small, of a vital resource can save the organism, then just maximizing the probability of its attainment would be the right response.

If the decision is not of urgent nature, the initial automatic preference is tentatively withheld, and is followed by an analytic processing (*System 2 or Type 2 processing,* (Stanovich & West, 2000; Stanovich 2011; cf. also Kahneman, 2011). The analytic processing system consists in comparing, weighting, and generally re-assessing the initial response of the automatic processing system. In fact, "One of the most critical functions of Type 2 processing is to override Type 1 processing" (Stanovich, 2011, p. 20). Indeed, the opportunity to win a greater amount, $x_1$, albeit with reduced probability, prompts the decision-maker to re-consider his initial automatic preference of prospect $(x_0, p_0)$, and analyze *all* available information, $x_0, p_0, x_1, p_1$. For this purpose, the decision-maker creates images of the possible consequences of each of the choices[1].

Our attempt to capture the analytic System 2 processing hinges on the assumption that prospect $(x_1, p_1)$ in effect *challenges* the chooser to abandon the *default prospect* $(x_0, p_0)$ and choose it, $(x_1, p_1)$, *the bold prospect*[2]. The challenge consists in facing the dissonance anticipated should the bold prospect be chosen, and the chooser loses the gamble. Psychologically, the dissonance anticipated may be of any of four kinds, or *modalities*: conative (regret, where the gambler would conclude he should have *acted* differently); cognitive (where the gambler realizes that the outcome conflicts with his prior *perceptions*, judgments or calculations); affective (feeling of disappointment and self-directed blame or anger); or valuative (degradation, shame or guilt upon the realization that one's choice-behavior conflicts with one's self-image, values, or adopted norms)[3]. An assessment of the *total* challenge presented by the various forms of dissonance is of interest in that, as we shall see, it can help predict people's inclination to choose the bolder prospects when faced with decision under risk; and possibly suggest which human traits correspond to higher inclination to prefer bold prospects. We turn therefore to an assessment of the magnitude of the described challenge. The following assessment is based on a reconstruction of possible mental processes that take place in the chooser's mind.

---

[1] Cf. "[better responses] come from processes of hypothetical reasoning and cognitive simulation that are a unique aspect of Type 2 processing." Stanovich, (2011, p. 22).
[2] The term *bold play* has been used in a somewhat similar sense in the study of subfair red and black casino with zero-one utility function (Dubins & Savage, 1965; Shye, 1968; Heath et al., 1972).

Faced with a binary gains problem, the chooser imagines the situation in which he would be, if he gambles on the bold prospect, $(x_1, p_1)$, and loses. His attention would then turn to the lower amount, $x_0$, which he could have had with a higher probability: *ceteris paribus*, the greater it is, the greater the anticipated dissonance; and therefore, the greater the challenge posed by the bold prospect. Hence, we conclude that this challenge *increases* with $x_0$.

The role of the greater amount, $x_1$, in this decision-making situation seems clear. It is this amount that lures the chooser and beckons her to pick the bold prospect. The chooser may contemplate: "should I lose my gamble on the bold prospect, I could rationalize that the prospect of winning this greater amount, $x_1$, justified my choice: I gave myself a chance". The greater is this amount, the lower the anticipated dissonance; and the lower the challenge presented by the bold prospect. We conclude therefore that this challenge *decreases* with $x_1$.

Finally, how do the given probabilities figure in the assessment of the challenge calculation? Having lost the bold gamble, the chooser would not like to feel he had made a grave error in judgment (cognitive dissonance). Thus, if $p_0$ is very close to $p_1$ the anticipated dissonance would be correspondingly small, since he could rationalize: "under these circumstances I could have just as well (with a similar probability) lost the smaller amount". But if $p_0$ is much higher than $p_1$, dissonances of all kinds could creep in. In general, then, the greater the difference $p_0-p_1$, the greater the anticipated dissonance; and the greater the challenge posed by the bold prospect. We conclude therefore that in general this challenge *increases* with $p_0-p_1$.

The Challenge Index (*CI*) in the case of gains, posed to the chooser by the bold prospect $(x_1, p_1)$ relative to the default prospect $(x_0, p_0)$, could therefore be given by:

(1) $$CI[(x_0, p_0), (x_1, p_1)] = \frac{f_0(x_0)}{f_1(x_1)} g(p_0 - p_1)$$

where each of $f_0$, $f_1$ and $g$ is a non-decreasing function of its respective argument.

Eq. (1) assumes that the subtraction of $p_1$ from $p_0$ is mentally performed on these same figures as presented to the decision-maker; and that the decision-maker subsequently transforms the difference $p_0-p_1$ by $g$. An alternative, perhaps more likely assumption would be that each of the probabilities $p_i$ is first transformed, or weighted, and *then* a mental assessment of the difference between the weighted probabilities is performed. This assumption results in a variant expression for CI:

(2) $$CI[(x_0, p_0), (x_1, p_1)] = \frac{f_0(x_0)}{f_1(x_1)}(w_0(p_0) - w_1(p_1))$$

where $f_0, f_1, w_0, w_1$ are non-decreasing functions of their respective arguments, and $w_0, w_1$ are such that $w_0(p_0) - w_1(p_1) > 0$.

---

[3] The first three behavior modalities are well known in psychological literature. The fourth, the valuative, concerns adopted beliefs and values that guide the individual (without the individual's actively evaluating them, hence it differs from the cognitive modality). The valuative modality can be shown by the Faceted Action System Theory (Shye, 1985; 2014) to complete a 2X2 cartesian product set, of which the first three modalities are a subset.

*2.2 The case of losses*

Let us look now at decision problems with $x_1 < x_0 < 0$. The decision-maker is faced with two options $(x_0, p_0)$ and $(x_1, p_1)$, with $p_1 < p_0$, each containing two numerical parameters, the (negative) outcomes, $x_i$, and the probabilities of receiving them, $p_i$. Again, the total information input consists of four numerical values to be processed mentally in order to reach a decision. But the decision-maker's initial impression of the problem is governed by his automatic mental processing system (*System 1 processing),* which, in this case too, is assumed to stress the probabilities rather than the outcomes. Hence, now, in the case of losses, we assume that a person's initial attraction, or *default* preference, would be $(x_1, p_1)$; that is, the prospect that promises the lowest probability for *any* loss. This assumption, too, may be regarded an extension of the certainty effect, where the preference of a sure avoidance of a loss is extended to the preference for a *surer* avoidance (smaller probability) of a loss. This assumption is supported by deeper considerations: Geared to avoid any negative consequence, the rapid System 1 processing tends to focus on the probabilities rather than on assessing the relative gravity of the consequences[4].

Time and circumstances allowing, the analytic System 2 processing re-examines the automatic preference of $(x_1, p_1)$, in view of the promise for a smaller loss (albeit with a higher probability) presented by the (loss-related-) *bold prospect* $(x_0, p_0)$, weighing now *all* four pieces of information, $x_0, p_0, x_1, p_1$. Again, as in the case of gains, the decision-maker is challenged to abandon the (loss-related-) *default prospect* $(x_1, p_1)$ and choose the *bold prospect*, $(x_0, p_0)$, instead[5]. The challenge consists in facing the dissonance anticipated should the bold prospect be chosen, and the chooser loses the gamble. We turn therefore to an assessment of the magnitude of this challenge. The

---

[4] Cf. in the context of crime prevention, "Research shows clearly that the chance of being caught is a vastly more effective deterrent than even draconian punishment" (National Institute of Justice, 2016). That is, increased risk of being caught (and getting *some* punishment) affects choice behavior more than increased punishments. See also Nagin, 2013.

[5] To illustrate the terminology and designations of this article consider the following examples of two problems:

|  | *Default Prospect* | *Bold Prospect* |
|---|---|---|
| Gain problem | (200, 0.8)<br>$(x_0, p_0)$ | (300, 0.6)<br>$(x_1, p_1)$ |
| Loss problem | (-300, 0.6)<br>$(x_1, p_1)$ | (-200, 0.8)<br>$(x_0, p_0)$ |

In both, gain and loss problems, prospect $(x_0, p_0)$ is characterized by the smaller amount (in absolute value) and by the higher probability, compared with the alternative prospect $(x_1, p_1)$. In gains, prospect $(x_0, p_0)$ is defined as the default; and in losses prospect $(x_1, p_1)$ is defined as the default. This designation allows for a single *CI* formula (3) for both gains and for losses. Note that in the case of losses, the definition of the bold prospect differs from the common definition of the *risky* prospect.

following assessment is based on a reconstruction of possible mental processes that take place in the chooser's mind.

Faced with a binary loss problem, the chooser imagines the situation in which he would be if he gambles on the bold prospect, $(x_0, p_0)$, and loses. His attention would then turn to comparing the lower probability, $p_1$, of the default (higher loss) prospect with the (higher) probability $p_0$, with which he just lost the gamble. If $p_0$ is very close to $p_1$ the anticipated dissonance would be correspondingly small, since the decision-maker could rationalize: "Under these circumstances I could have just as well (with a similar probability) lost the higher amount, $x_1$." But if $p_0$ is much higher than $p_1$, the dissonance could be great, since had he chosen the default $(x_1, p_1)$ he could have, with greater probability, avoided *any* loss. In general, then, the greater the difference $p_0-p_1$, the greater the anticipated dissonance; and the greater the challenge posed by the bold prospect. We conclude therefore that in general this challenge *increases* with $p_0-p_1$.

The appeal, if any, of the bold prospect lies in its lower loss, $|x_0|$, ($|x_0|<|x_1|$). *Ceteris paribus*, the greater this lower loss, $|x_0|$, the lower its appeal and the greater the anticipated dissonance; and therefore, the greater the challenge posed by the bold prospect. Hence, we conclude that this challenge *increases* with $|x_0|$.

The role of the greater loss, $|x_1|$, in this decision-making situation, seems clear. It is this larger loss that may tempt the chooser to consider the bold alternative with its promise for a lower loss. The higher the default loss, $|x_1|$, the lower the anticipated dissonance; and the lower the challenge presented by the bold prospect. We conclude therefore that this challenge *decreases* with $|x_1|$.

Adopting the assumption that it is the difference between *weighted* probabilities (rather than the weight of the difference between the probabilities) that should incorporated into the expression for the Challenge Index, *CI*, eq. 2 holds for both loss problems as well as for gains problems:

(3) $$CI[(x_0, p_0), (x_1, p_1)] = \frac{f_0|x_0|}{f_1|x_1|}(w_0(p_0) - w_1(p_1))$$

Where $f_0, f_1, w_0, w_1$ are non-decreasing functions of their respective arguments, and $w_0, w_1$ are such that $w_0(p_0) - w_1(p_1) > 0$.

### 3. The Challenge Theory of Choice under Risk

Our main purpose is to explain and facilitate the prediction of people's inclination to prefer the bold prospect in a specified binary decision problem, where both prospects involve monetary gains (or, alternatively, both involve monetary losses) with known probabilities. Here, *preference inclination* is interpreted for a given population as the proportion of persons in that population who choose the bold prospect in the specified problem. The notion of Challenge is developed, and the Challenge index (*CI*) formulated, with this purpose in mind.

*3.1 The General Challenge Hypothesis*: The greater the challenge, *CI*, associated with a binary simple problem, the less likely are people to choose the bold prospect.

*Rationale:* The challenge as defined presents an obstacle for the chooser in deciding to prefer the bold prospect over the default. The greater the challenge, the fewer the people who would surmount this obstacle.

*Scales:* To test the general hypothesis, further specifications need to be made concerning the functional forms of $f_0, f_1, w_0, w_1$. For examining the viability of the psychologically based Challenge Theory as a descriptive theory, we adopt here those widely used forms that seem to us both promising and simple (not involving too many free parameters). Thus, for $f_0$ and $f_1$ we adopt the power function:

(4) $$f_0 = x^{a_0} \qquad f_1 = x^{a_1}$$

and for $w_0, w_1$, we adopt the two-parameter "linear in log odds" functional form advanced by Gonzalez & Wu (1999):

(5) $$w_i(p) = \frac{\delta_i p^{\gamma_i}}{\delta_i p^{\gamma_i} + (1-p)^{\gamma_i}} \qquad i = 0, 1$$

characterized by two parameters: $\gamma_i$, that determines mainly the curvature of the weighting function (ranging from near linear to near step-function); and $\delta_i$, that determines mainly the elevation of the weighting function, i.e., the overall inclination to underweight or overweight probabilities.

As a measure of association between *CI* and the proportion of those preferring the bold prospect, we use the correlation coefficient, *r*.

With the present choices of functional forms for the outcomes and the probabilities, Challenge Theory allows, in principle, for up to six free parameters for gains: $a_0, \gamma_0, \delta_0, a_1, \gamma_1, \delta_1$ and up to six free parameters for losses: $a'_0, \gamma'_0, \delta'_0, a'_1, \gamma'_1, \delta'_1$. Naturally, the greater the number of free parameters, the better the fit of the model to the data of a particular study, but the greater the risk of overfitting. In this study we examined 3, 4 and 6 parameter models and chose to present and elaborate on the model with the 4 parameters: $a_0, a_1, \gamma, \delta$ (i.e., $\gamma_0 = \gamma_1$ ; $\delta_0 = \delta_1$) for gains; and with the 4 parameters: $a'_0, a'_1, \gamma', \delta'$ for losses. This choice was based primarily on the extent of improvement in *r*, the resultant correlation coefficient between *CI*, the challenge level, and $P_b$, the proportion of respondents preferring the bold option. (See discussion in Section 5.5 below.)

### 3.2 Testing the General Challenge Hypothesis

#### 3.2.1 Data

The data set for this study consists of responses to a questionnaire obtained from a sample of 126 twelfth-grade students at two upscale high schools in Jerusalem. Students in these schools are generally considered to be conscientious and highly motivated and they seemed to address themselves to the task of completing the questionnaires with due attention and concentration. No material incentives were offered.[6]

The questionnaire included 44 simple binary choice problems, 22 gain problems and 22 loss problems. Problems were selected to represent a fairly wide range, with gains and losses ranging from 30 to 9000 Israeli Shekels; and probabilities ranging from 0.01 to 1.

Analyses were performed also on data collected from samples of students and faculty members by Kahneman and Tversky (1979). Eleven simple binary choice problems were found in these data (six gain problems and five loss problems).

For every choice problem, $v_i$, two variables were recorded:

i. The Challenge Index, $CI(v_i)$, as computed by equation (3);
ii. The percentage, $P_b(v_i)$, of respondents who choose the bold prospect.

To test the Challenge Hypothesis, Pearson correlation coefficient, $r$, between $CI$ and $P_b$ was determined while searching for optimal values for the parameters $x_0, x_1, \gamma, \delta$ defined in Eqs. (4) and (5). The optimization was performed using Microsoft Excel GRG solver.

In accordance with the hypothesis, we expected high *negative* correlations between $CI(v_i)$ and $P_b(v_i)$.

*3.2.2 Results*

Analyses were performed separately for gain problems and for loss problems since evidence points to the fact that these two classes of problems are perceived differently by decision makers (e.g., Kahneman & Tversky 1979; Tversky & Kahneman, 1992. See also discussion below). Hence the values of parameters, $a_0, a_1, \gamma, \delta,$ may well be expected to differ for losses and for gains.

Results obtained support the general challenge hypothesis: the greater the Challenge Index, *CI,* associated with a binary decision problem, the smaller the percentage of respondents who chose the bold prospect. This is found for the set of gain problems ($r = -.92$) as well as for the set of loss problems ($r = -.93$). See Table 1.

---

[6] Incentives could interfere with the main purpose of the study, namely, to understand decisions made by unschooled people as well as by schooled people in their everyday life, decisions that (according to CT) result from of the mental interaction between the (automatic) system 1 and the (deliberate, calculating) system 2. Incentives are known to encourage the deliberate, calculating system 2. (Often this is indeed why they are used.) Since most of our respondents are familiar with the notion of *expected value,* they would regard the questions as a test of their knowledge (and hence aim for the economically "correct" answer), rather than reveal their endogenous inclination. Thus, the spontaneous balance between the two processing systems could be impaired. In the context of descriptive theories in general (not just dual system theories), some writers believe that incentives are not essential. For example, Camerer (1989) finds that "Subjects who actually played a gamble were no more reliable than subjects who did not play" and concludes his test of the effect of incentives: "Incentives make little difference in subjects' choices among gambles involving gains" (pp. 82-83). Tversky & Kahneman (1992) conclude their discussion on the question of incentives thus: "The similarity between the results obtained with and without monetary incentives in choice between simple prospects provides no special reason for skepticism about experiments without contingent payment" (p. 316).

**Table 1. Pearson Correlation Coefficients, *r*, between Challenge Index, *CI*, and Observed Proportion, $P_b$, of Bold Responses for Gains and for Loss problems**

|  | *r* | 0.95 confidence interval | $a_0$ | $a_1$ | $\gamma$ | $\delta$ |
|---|---|---|---|---|---|---|
| Gains Problems *n*=22; *N*=126 | **-.919** | (-.966, -.813) | 1.1936 | 1.2285 | 0.7336 | 2.6245 |
| Loss Problems *n*=22; *N*=126 | **-.931** | (-.971, -.839) | 1.3349 | 1.4337 | 0.6505 | 3.5565 |

Interestingly $a_0$, $a_1$ >1 both for losses and for gains, suggesting that, under the conceptual framework developed here, a convex value function optimizes predictions of bold choices. Indeed, the possibility of convex value (utility) functions, or functions are convex in segments of the monetary domain have been proposed and studied in the literature (e.g., Dubins & Savage, 1965; Friedman & Savage, 1948; Markowitz, 1952;). However, CT proposes a reformulation of the question of outcome values. See Subsection 5.4 of the discussion. As for the weighting functions, they were found to adhere to the expected shape -- concave at the lower probability range and convex at the higher part. But they switch their shape at a very high point on the 0-1 probability scale. Again, a different perspective on this question is suggested in Subsections 5.4 of the discussion.

The results obtained here for the General Challenge Hypothesis lends support to the psychological assumptions and considerations of the Challenge Theory.

**Figure 1. Scatter Diagrams and Linear Regression Lines of $P_b(v_i)$ vs. $CI(v_i)$ (a) for the 22 Gains Problems and (b) the 22 Loss problems**

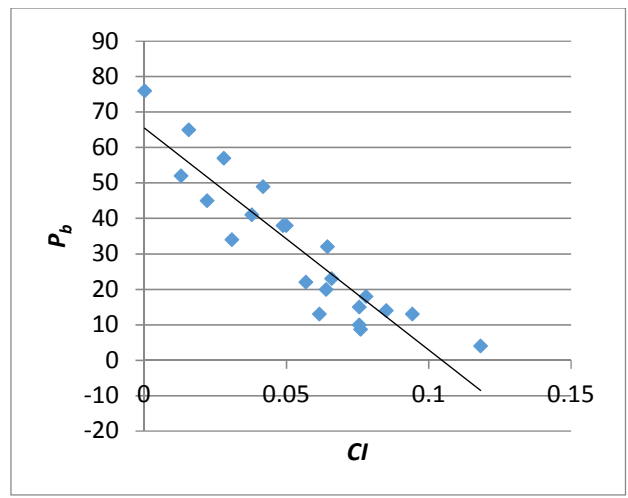

**(a) Gains**

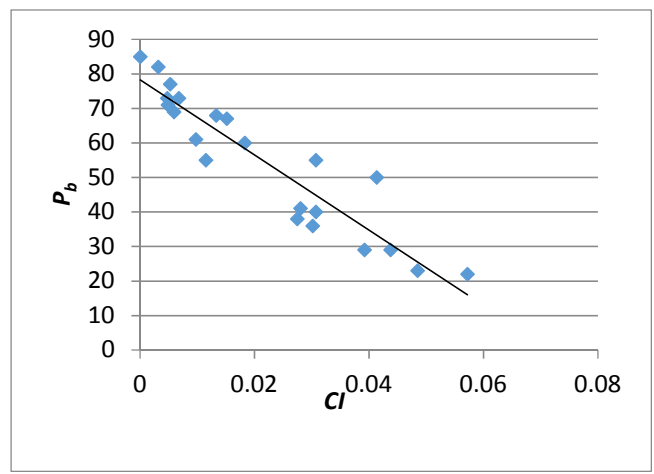

**(b) Losses**

Of course, if the entire sample of 44 gain and loss problems is processed together (i.e., imposing the same set of four parameters on both, gain and loss problems), a lower correlation ($r=-.87$) is found between $CI$ and $P_b$. See first row in Table 2. The second row in Table 2 shows a very high correlation ($r=-.99$) for Kahneman & Tversky's (1979) sample of 11 gain and loss problems. However, this sample of problems is small, and the $CI$ values are not well spread over their range.

**Table 2. Pearson Correlation Coefficients, *r*, between Challenge Index, *CI*, (Eq. 3) and Observed Proportion, $P_b$, of Bold Responses for All (Gain & Loss) Problems of This Study and of Kahneman & Tversky's (1979) Study**[*]

|  | *r* | 0.95 confidence interval | $a_0$ | $a_1$ | $\gamma$ | $\delta$ |
|---|---|---|---|---|---|---|
| All Problems of this study $n$=44; $N$=126 | **-.877** | (-.929, -.780) | 1.5392 | 1.4806 | 0.7633 | 2.9011 |
| All Binary Problems of KT (1979) study $n$=10; $N$~70 | **-.989** | (-.997, -.956) | 3.0 | 0.6145 | 0.5599 | 2.7184 |

[*] Note that in Kahneman & Tversky's (1979) study, the expected value of the bold prospect is greater or equal to that of the default prospect, in all problems. In this study (where the expected value function plays no role) no such constraint was imposed on the selection of problems.

**Figure 2. Scatter Diagrams and Linear Regression Lines of *P(v_i)* vs. *CI(v_i)* (a) for all 44 (Gain & Loss) Problems of This Study and (b) for all 11 (Gain & Loss) Problems of Kahneman & Tversky's (1979) Study**

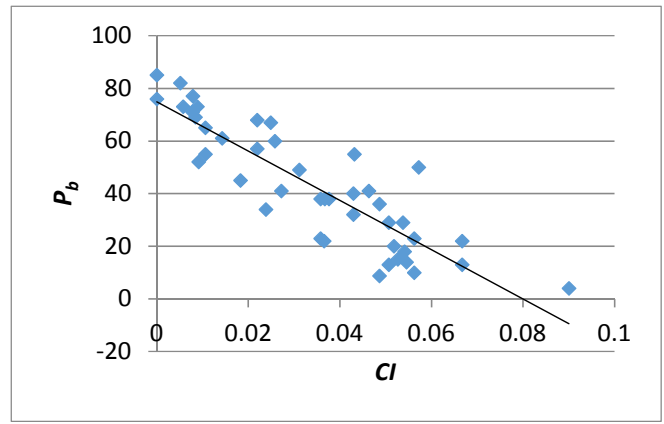

**(a) Present Study**

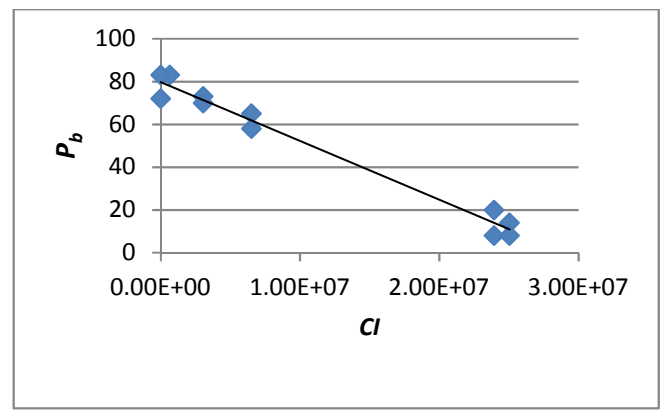

**(b) Kahneman & Tversky, 1979**

Overall the results, demonstrating a very high negative and linear relationship between *CI* and the proportions of bold choices, provide substantial evidence for the proposed Challenge Theory. A two-fold cross validation performed on the data (Appendix A) provides further support for the proposed model. Nevertheless, replications must follow to establish the theory and examine whether different kinds of populations may require somewhat different optimization parameters to attain the high correlations hypothesized.

**4. Determinants of Bold Decisions**

As a descriptive, psychologically oriented theory, the Challenge Theory of decision making opens the way to investigate individual differences that are associated with decision under risk behavior. As an illustration of this possibility we formulate and test the following hypothesis.

*4.1 The Resource Hypothesis:* The more personal resources a person has, the more inclined that person would be to choose the bold prospect.

*Rationale:* People with more reserve resources such as money, relevant skills, or self-confidence are better equipped to meet the challenge presented by the bold prospect; they are more likely to be able to contain and live with the kinds of dissonance anticipated in case they lose their bold gamble.

*4.2 Testing the Resource Hypothesis*

Two of the external (background) variables in the questionnaire could be interpreted as related to personal resources and may serve as indicators of that trait. Hence, they were hypothesized to be associated with the inclination to choose the bold prospect in the simple binary choice problems presented. These variables are:

- *Gender,* where male participants are assumed to have somewhat greater access to personal resources (e.g., financial skills).

  That men are more inclined take risks in financial matters than women, is well documented (e.g., Powell & Ansic, 1997; Charnessa & Gneezy, 2012; Barber & Odean, 2001). The reasons behind such findings are often traced to social or psychological factors (social status, self-confidence). Here we propose to attribute this difference in risk attitude to men's enhanced familiarity and skills in financial matters, which conceptually fall under the general umbrella of enhanced resources.

- *Personal Earnings*, where those with greater earning power are assumed to have more personal resources than others.

  People with greater earning power feel they can recover more easily from the dissonances expected following an unsuccessful gamble, than those with smaller earning power.

  For the present purpose, the variable chosen for assessing individual's earning power was student's own earnings (not their family incomes, which could be used as another, distinct indicator). The tacit assumption was that the actual experience of personally earning money suggests some financial skill (considered here as a kind of resource) that would be reflected in respondents' attitudes and gambling behavior, as hypothesized. Hence, participants were asked what their hourly pay was in their recent job, if any. Nearly eighty percent (78.57%) of the responded with a figure higher than 0.

To test the Resource Hypothesis, we first defined a *gain bold player* as one whose number of bold choices from amongst the 22 gains problems exceeds the average number (6.94) of bold choices in the sample. Similarly, we define a *loss bold player* as one whose number of bold choices from amongst the 22 loss problems exceeds the average number (11.79) of bold choices in the sample. Then the proportions of gain bold players and of loss bold players were computed for each subgroup— for males and females; and for "rich" and "poor" (defined according to whether the pay per hour exceeded the median or not). The results are shown in Table 3.

**Table 3. Proportions (%) of Gains-Bold-Players and of Loss-Bold-Players
(i) by gender; (ii) by Earnings**

| (i) Gender | Male | Female | Difference | Significance |
|---|---|---|---|---|
| Gains | 65.5 | 49.3 | 16.2 | 0.07 |
| Losses | 60.0 | 62.0 | -2.0 | n.s. |

| (ii) Earnings | Rich | Poor | Difference | Significance |
|---|---|---|---|---|
| Gains | 65.1 | 47.6 | 17.5 | 0.05 |
| Losses | 60.3 | 61.9 | -1.6 | n.s. |

Table 3 supports the Resource Hypothesis only for gain problems. Men more than women, and separately, the "rich" more than the "poor", were inclined to face the challenge posed by the bold prospects of gain problems. However, no significant differences were found by gender or by earnings with respect to loss-problems. This interesting result, at variance with our hypothesis, calls attention to the essential asymmetry between losses and gains. We shall comment on this asymmetry below, in the discussion section.

## 5. Discussion

*5.1 Cutting the umbilical cord*

It is only natural that in the search for descriptive theories of choice under risk, prescriptive EUT has been adopted the as a point of departure: EUT is based on the plausible axioms of rational choice and on the compelling von Neumann-Morgenstern utility theorem. But, in the face of systematic violations of EUT observed in human behavior, descriptive theories, including the psychological-process theories, refined and improved EUT with important insights from psychology. The Challenge Theory of decision under risk proposed in this article aims to complete this trend by asserting that the psychological approach to decision under risk can stand on its own, without the historic normative economic platform.

As a psychological theory, Challenge Theory seeks to trace the mental processes that take place in decision making and specifically with reference the possible shift from a default choice to the alternative. This is in contrast with the economic approach which assumes that people evaluate and compare available prospects as separate, distinct entities. In this sense, CT is a processual rather than a static theory. Moreover, rather than identify the one, universally best decision, CT takes account of individual differences in preferring the bold prospect (as defined here) and enables relaing such preferences to other individual characteristics.

CT presented in this study points to a research approach that aims to predict risk behavior and tests this approach in the case of simple binary decision problems. The viability of the proposed approach must be evaluated by suitable experimental replications and by studying the values of the optimizing parameters in diverse populations and experimental circumstances.

*5.2 The psychological model*

CT considers decision-makers' anticipated feelings towards the possible outcome of the prospect not chosen. However, CT does not incorporate these feelings (of regret or rejoicing) into the prospects offered to obtain a value for each prospect. Rather, having specified the default prospect preferred by the "automatic mind" (System 1 Processing), it assesses the difficulty, or the *challenge* experienced by the decision maker in abandoning this initial preference in favor of an alternative, the *bold* prospect. The decision maker's task amounts then to deciding whether to meet this challenge level. The decision itself is personal and is determined by personality characteristics and circumstances to be explored.

Challenge Theory posits that the decision process in binary risk problems involves comparing the two outcomes, $x_0$, and $x_1$, and proceeds to represent this comparison in the form of a ratio between the monotonically transformed outcomes. This ratio constitutes one factor in the *CI* equation. It is plotted in Figure 3 as a function of its argument-pair ($x_0$, $x_1$). CT decision process also involves comparing the two probabilities, $p_0$ and $p_1$, associated with the outcomes. The result of this comparison, in the form of the difference between the monotonically transformed probabilities, constitutes the second factor in the *CI* equation. This difference is plotted in Figure 4 as a function of its argument-pair ($p_0$, $p_1$). Thus, CT decision process contrasts with that implied by most theories where options ($x_0$, $p_0$) and ($x_1$, $p_1$) are first evaluated separately to obtain a total worth for each, and then the two options are compared, typically by computing the difference between their total worth or an increasing function of that difference (e.g., Erev et al.*,* 2010).

*5.3 Effects: certainty, reflection, low probabilities and loss aversion in CT perspective*

It is of interest to examine the manifestations in CT of effects proposed by Kahneman & Tversky (1979) to account for observed choice behavior. Such an examination can cast light and enhance understanding of the paradigmatic shift embodied in CT. Table 4 illustrates effects that are apparent in binary non-mixed gambles.

**Table 4. Illustration of Effects and their Manifestation in Terms of *CI***

| Effect | Problem D=Default **B=Bold option** | % Bold | *CI*\*100 |
|---|---|---|---|
| 1. Certainty/Allais Paradox Kahneman & Tversky, 1979 following Allais, 1953. | D: 3000, 1 **B: 4000, .80** | 32% | 6.64 |
| | D': 3000, .25 **B': 4000, .20** | 57% | 2.80 |
| 2. Reflection Kahneman & Tversky, 1979 | D: 3000, 1 **B: 4000, .80** | 32% | 6.64 |
| | D': -4000, .80 **B': -3000, 1** | 40%* | 3.08 |
| 3. Overweighting low probabilities | D: 3000, .02 **B: 6000, .01**\*\* | 65% | 1.57 |
| 4. Loss Aversion | See Table 5 and discussion below | | |

\* Recall that in loss problems, the often-labelled "safer" option (i.e., the option of higher probability of a smaller loss; here, (-3000, 1)) is defined in CT as the *bold* option and the 'risky' option is defined in CT as the *default* option.

\*\* In Kahneman & Tversky (1979) the problem was: (3000, .002; 6000, .001).

The *certainty effect* is illustrated by the first problem of Table 4, row 1, (3000,.8; 3000,1). Respondents' preferences in this problem are affected by the relatively high *CI* (*CI*\*100=6.64) which according to CT central hypothesis entails a relatively low proportion of bold choices (32% in the present study; 20% in Kahneman & Tversky, (1979)). The two choice-problems in row 1 illustrate a two-outcome simplified version of Allais paradox (Kahneman & Tversky, 1979), where the probabilities in the first problem were multiplied by 0.25 to obtain those of the second problem. The seemingly inconsistent high preference rates for the less certain option (B') in the second problem in row 1, (3000, .25; 4000, 20), is explained by CT simply: Because of the smaller difference between the two probabilities (0.25-0.20=0.05), *CI* is much lower here (*CI*\*100=2.80) than in the first problem (6.64), prompting the gambler to go for the bold option.

The *reflection effect* is expressed in CT simply as adherence (by the majority) to the respective default option: (3000, 1) in gains and to (-4000, .80) in losses. See problems in Table 4 row 2, where a minority (40%) chose bold (in Kahneman & Tversky (1979) the corresponding figure was18%). While in Prospect Theory the reflection effect is regarded as a curious reversal -- a surprising switch from risk aversion to risk seeking (Kahneman & Tversky, 1979 p. 268); In CT the reflection effect is built-in into the conceptual basis of the theory.

The *effect of overweighting low probabilities* is illustrated in the problem in Table 4 row 3. In CT terms this overweighting is a consequence of the very small difference (0.01) between the probabilities which results in low *CI* (*CI*\*100=1.57), i.e., strong inclination to shift from the default to the bold option, (6000, 0.01). An essentially different effect, also considered by Kahneman & Tversky (1979, p.281) as an instance of overweighting low probabilities, occurs in the problem (5, 1; 5000, 0.001), where only *one* probability is low and 72% were found to prefer the bold option (5000, 0.001). Kahneman & Tversky's explanation involves the shapes of the value and weighting functions. But CT explains the observed preference differently, namely, by the very small ratio of $x_0^{a_0}/x_1^{a_1}$ resulting in an extremely low *CI*. These arguments hold equally for the corresponding loss problems (Problems 8' and 14' in Kahneman & Tversky, 1979).

The predominance of loss-avoidance over gain-seeking (a phenomenon known as *loss aversion*) is manifested by CT in the difference between $CI^+$, the *CI* of a given gain problem, and $CI^-$, the *CI* of its symmetrical (mirror-image) loss problem[7]. Table 5 presents this difference $(CI^+ - CI^-)*100$ for the five symmetrical gain/loss problem-pairs in our study. See last column in Table 5. The essential thing to notice about these differences is their *sign*: they are all positive, i.e., $CI^+ > CI^-$. According to CT this means that people are more likely to switch from the default option to the bold option in losses than in gains; hence, more likely to avoid the greater loss than to opt for the greater gain of a similar size. The *magnitudes* of the recorded differences ($CI^+ - CI^-$) may well reflect the *extent* of loss aversion in each case, suggesting that loss aversion depends on the probabilities as well as on the outcomes.

It is noteworthy that the four effects discussed above are explained in CT by the single parameter *CI*, not by reference to the shapes of value functions or weighting functions, whose stability is not assumed by CT.

**Table 5. CT Representation of Loss Aversion:**

***CI* in a Gain Problem is Greater than in the Symmetrical Loss Problem**

| Illustration No. | Problem-Pair D=Default **B=Bold option** | % Bold (This Study) | *CI*\*100 | Δ*CI*\*100 ($CI^+ - CI^-$) \*100 |
|---|---|---|---|---|
| 1 | D: 3000, 1 **B: 4000, .80** | 32% | 6.64 | 3.36 |
| | D': -4000, .80 **B': -3000, 1** | 40% | 3.08 | |
| 2 | D: 3000, .25 **B: 4000, .20** | 57% | 2.80 | 1.47 |
| | D': -4000, .20 **B': -3000, 25** | 68% | 1.33 | |
| 3 | D: 3000, .90 **B: 6000, .45** | 9% | 7.61 | 4.59 |
| | D': -6000, .45 **B': -3000, 90** | 36% | 3.01 | |
| 4 | D: 240, 1 **B: 1000, .25** | 23% | 6.59 | 3.84 |
| | D': -1000, .25 **B': -240, 1** | 38% | 2.74 | |
| 5 | D: 3000, .02 **B: 6000, .01** | 65% | 1.57 | 0.41 |
| | D': -6000, .01 **B': -3000, .02** | 55% | 1.15 | |

---

[7] I.e., the loss problem created from the given gain problem by switching the outcome signs from positive to negative.

Finally, it is interesting to comment on how CT conceptual framework can take account of dominated prospects directly and simply: If $x_i > x_j > 0$ and $p_i > p_j$ the analytic mind (System 2 processing) reinforces and readily confirms the automatic mind initial evaluation based on the *p*'s. No prior editing phase (suggested in Kahneman & Tversky, 1979) is required; the entire decision process is naturally captured by the psychological model.

*5.4 The impact of outcomes and probabilities on decisions: CT vs. EUT-based theories*

EUT-based descriptive theories use value and weighting functions to transform outcomes and probabilities. The transformed quantities (often interpreted as subjective outcomes and subjective probabilities, respectively) serve to compute the total worth of each prospect for explaining empirically observed preferences. In CT, an outcome, $x_0$ or $x_1$, is not conceived as impacting on the decision by itself, but rather as acting within the pair ($x_0, x_1$), where the outcomes are transformed and combined to form the ratio, $x^{a_0}/x^{a_1}$, constituting the first factor in the *CI* equation. Hence, it is the impact of this *pair* on decision (in effect, on the value of the challenge index, *CI*) that is of interest. This is illustrated in Figure 3. Similarly, the transformed probabilities are combined to form the difference, $w_0(p_0) - w_1(p_1)$, constituting the second factor in the *CI* equation illustrated in Figure 4.

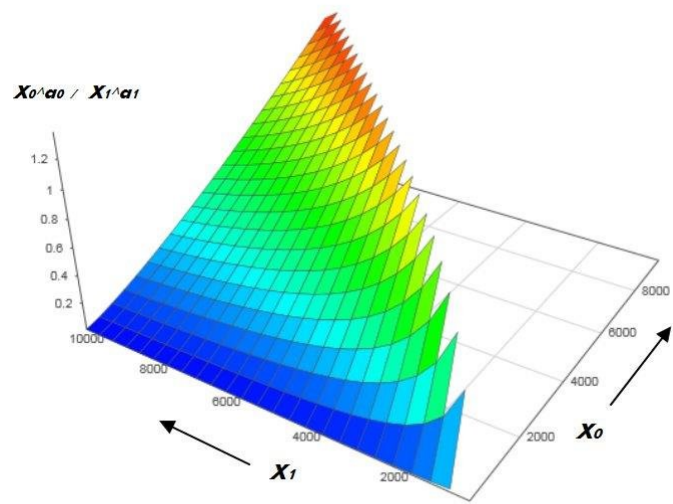

**Figure 3. The First Factor, $x_0{\wedge}a_0 / x_1{\wedge}a_1$, of *CI* as a Function of the Two Outcomes: Illustration with $a_0$=1.19, $a_1$=1.23**

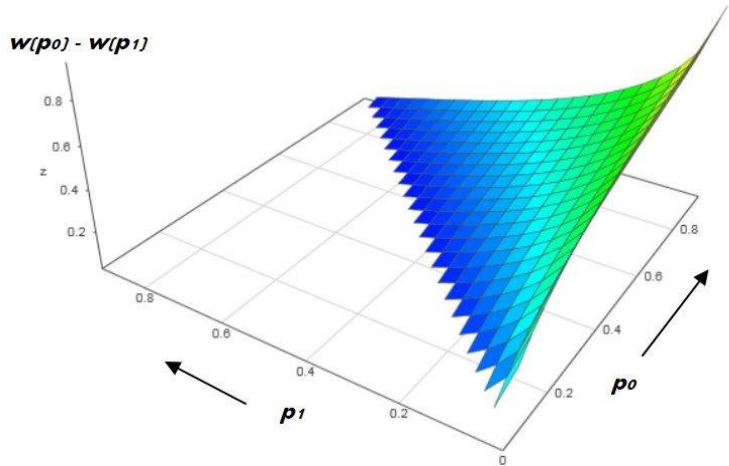

**Figure 4. The Second Factor, *w(p₀)-w(p₁)*, of *CI* as a Function of the Two Probabilities, where** $w(p) = \frac{\delta p^\gamma}{\delta p^\gamma + (1-p)^\gamma}$ **. Illustration** with *γ*=0.73; *δ*=2.62

Observed decision are now explained by the single quantity *CI*, not by shapes of value functions or weighting functions, as such. And since each of the outcomes offered is not considered separately (but only in conjunction with the other outcome in the pair), the transformed outcome need not be interpreted as representing a subjective value of that outcome.

Yet, it is of interest to note the magnitudes of the extracted parameters, $a_0, a_1, \gamma,$ and $\delta$, which jointly optimize predictions, and compare them with those that may be obtained in future studies. Do they vary with the type of population? Do they vary with the magnitudes of the amounts or the probabilities? Or with the spacings between the amounts or the probabilities? And, ultimately, how good are the resultant predictions in cross-validation exercises?

*5.5 How many free parameters?*

As noted above (Section 3.1), this paper details results obtained for the choice of four free parameters. This choice was made after comparing results obtained as the number of parameters increases from 3 to 4 and from 4 to 6. We found that while in gains the improvements in *r* were small (from -.9145 to -.9174 to -.9276); in the case of losses a larger improvement (from -.9098 to -.9314) was recorded when the number of parameters increased from 3 to 4. (But in the 6-parameter model improvement was small again, with *r*= -.9376.) Hence for losses we opted for the 4-parameter model and for reasons of symmetry adopted the same number of parameters for gains. However, our evidence for the optimal number of free parameters is not conclusive and the simpler model of three parameters, *a, γ, δ* (i.e. $a_0=a_1$), for gains and three, *a', γ', δ',* (i.e. $a'_0=a'_1$), for losses which yielded here a good fit with the data (*r*=-.9145 for gains and *r*=-.9098 for losses), should be re-examined in future studies. At this point it is worth noting that the technical simplicity of the 3-parameter model in each case (gains and losses) is accompanied by an appealing substantive-interpretative advantage: With the outcomes raised to the same power, the ratio $x_0/x_1$ becomes a basic, psychologically meaningful quantity.

In conclusion of this subsection, we note that alternative functional forms have not been systematically tested in this study. The power functional form adopted here for transforming the outcomes has been justified and employed by Tversky and Kahneman (1992) and by others (e.g. Harrison et al., 2009). As for the functional form for transforming probabilities, we did consider the one-parameter functional form $w(p) = \frac{p^\gamma}{(p^\gamma + (1-p)^\gamma)^{1/\gamma}}$ employed by Tversky and Kahneman (1992) but found it to produce much poorer results in this study ($r$=-0.71 for gains and -0.60 for losses) than the function employed here. For comparison, when no parameters at all were used, (i.e., if $CI = \frac{|x_0|}{|x_1|}(p_0 - p_1)$) The results were about the same for gains ($r$=-0.69) but lower for losses ($r$=-0.47).

*5.6 The role of extant wealth: three views*

What is the role of a person's existing wealth in preferring one prospect over another when facing a decision problem? In EUT the objects of choice are probability distributions over *total* wealth. Hence, the expected consequences of the decision (gains or losses) are each combined with extant wealth to assess the value to the decision maker of the resultant total wealth. Prospect Theory, based on people's observed behavior, finds that existing wealth usually marks a reference point relative to which, gains and losses are defined and assessed by the decision maker. But PT assigns no essential role to existing wealth in decision making itself and reveals that "The carriers of values are gains and losses, not final assets" (Tversky & Kahneman, 1992, p. 299). Challenge Theory, in its reconstruction of the decision process, essentially accepts this view. However, by foregoing the need to determine a normative choice and allowing instead individual differences, CT opens the door for investigating associations between risk behavior and other individual characteristics, such as wealth, as was demonstrated above in Section 4, thereby assigning a role to extant wealth. Moreover, CT allows for associating risk behavior with *flexible* interpretations of wealth such as those that include present or future earning power, confidence in future life-prospects and other manifestations of individual resources.

*5.7 The gain-loss asymmetry*

Arithmetically, positive and negative numbers are symmetrical about the 0 point. Psychologically, the prospect of obtaining a positive amount is interpreted and processed by the mind differently from incurring a negative amount, i.e., a loss. This psychological asymmetry is manifested for example by the phenomenon labeled "loss aversion" by Tversky & Kahneman (1992). The reasons for this asymmetry could be as follows.

To conceive the loss of an amount of money, one must first picture that amount thereby endowing it with a sort of mental existence. Then one must perform the opposite mental operation of annulment to conceive the absence of that amount[8]. This operation which involves an extra mental burden in conceiving a loss—compared with that of conceiving a gain -- may account for asymmetries between gains and losses in choice behavior. For the extra cognitive complexity of conceiving a loss (a difficulty that may be coded by the automatic processing system as an added risk) inclines the decision maker towards the less risky option. This is a psychological-*processual* explanation, referring to the mental processing of input information.

---

[8] Cf. *Ironic Process Theory*; e.g., "Don't think of a white bear".

But there is deeper, more *substantive* asymmetry between gains and losses. A gain is perceived by the living organism as an incidental favorable event, one whose recurrence is likely to be advantageous. A loss, on the other hand, is interpreted by the organism as a potential threat to its very structure; since *its* recurrence can destroy the organism as a living system.[9]

Challenge Theory exhibits both, similarities and differences, between gains and losses. A basic formal similarity is embodied in Eq. 3, a single formula that assesses the challenge-level for both gains and losses. This is noteworthy because our reconstruction of the cognitive decision process involving gains and that involving losses, were conducted separately and independently.

An interesting difference between gain and loss problems appears in our investigation of the association between bold choice behavior and personal resources: Thus, while as hypothesized, people who earn more are more likely to prefer the bold prospect in gain problems, no such association was found in the case of loss problems (Table 3). It seems that when losses are in sight the poor are just as likely to prefer the bold prospect as the rich precisely because (pursuing System 2 processing) that prospect sets a limit on the amount of possible loss, protecting them from excessive loss, to which they are more vulnerable.

Finally, the present findings exhibit similarity in the order-of-magnitude of the four key parameters, $a_0, a_1, \gamma, \delta$, inferred for gain problems and inferred for loss problems. The differences originate in the psychological difference between gains and losses and, as argued above, represent the predominance of loss-avoidance over gain-seeking behavior, known as *loss aversion*.

*5.8 Conclusion*

Challenge Theory tackles the question of explaining choice under risk from a psychological perspective rather than an economic one. Forgoing the relics of EUT, such as the rationality axioms, the notions of expected value, risk-seeking or risk-aversion and the concern with identifying the *one* most preferred choice, Challenge Theory suggests a new paradigm and proposes a simple formula, based on a reconstruction of mental decision processes, for describing the relative frequency of people who prefer one prospect over another in a class of simple binary choice problems. The possibility that opens up for associating individual choice behavior with other personality traits carries with it the promise of enriching the study of behavior under risk, with possible relevance to domains of application.

---

[9] Excessive gains can also be detrimental to a system, constituting a potential threat to its structural stability (Shye, 1989; 2014). However, usually they are not perceived as a threat because of a tacit assumption that resources are generally limited, rendering the occurrence of harmful gains unlikely.

## Appendix A. Two-Fold Cross Validation

### 1. Procedure

The sample of 126 respondents was randomly divided into two subsamples, A and B. The following cross-validation procedure was carried out separately for gain-problems and for loss problems. Each subsample in its turn served as a *Training Subsample* on which the optimization procedure was performed, resulting in

(i)   the optimized $r(CI, P_b)$, the correlation between Challenge Index and the proportion choosing the bold prospect; and in

(ii)  four optimizing value-function and weighting-function parameters.

Then, these four parameters were employed to compute $r(CI, P_b)$ in the alternative subsample, *Testing-Subsample*.

### 2. Results

Results of this cross-validation exercise are satisfactory both in the case of gains and in the case of losses: $r(CI, P_b)$ remain very high in all testing-subsamples. The correlations obtained in the testing subsamples are as high as in the training subsample in the case of gains, and almost as high in the case of losses.

**Table 4. Results of Two-Fold Cross Validation**

In this table, the row "A => B" indicates results of the cross validation procedure where A is the training subsample and B is the testing subsample. In "B => A" subsample roles are reversed

|  | Training-Subsample Correlation | Value Function Parameters Exponents | | Weighting Function Parameters Curvature Elevation | | Testing-Subsample Correlation |
|---|---|---|---|---|---|---|
|  | $r$ | $a_0$ | $a_1$ | $\gamma$ | $\delta$ | $r$ |
| **GAINS** | | | | | | |
| A => B | -0.9 | 1.18 | 1.15 | 0.7 | 2.41 | -0.9 |
| B => A | -0.92 | 1.27 | 1.23 | 0.77 | 2.84 | -0.92 |
| Average | -0.91 | 1.225 | 1.19 | 0.735 | 2.625 | -0.91 |
| **LOSSES** | | | | | | |
| A => B | -0.89 | 1.74 | 1.48 | 0.65 | 2.72 | -0.88 |
| B => A | -0.91 | 1.39 | 1.33 | 0.67 | 3.66 | -0.89 |
| Average | -0.9 | 1.565 | 1.405 | 0.66 | 3.19 | -0.885 |